\documentclass[ACS,STIX2COL]{WileyNJD-v2}
\articletype{Research Article}%

\begin{document}

\title{Digital resolution enhancement in low transverse sampling optical coherence tomography angiography using deep learning}

\author[1,4,5]{Ting Zhou}
\author[1,2,5]{Kang Zhou}
\author[1]{Jianlong Yang*}
\author[1]{Liyang Fang}
\author[3]{Yan Hu}
\author[1]{Jun Cheng}
\author[1]{Yitian Zhao}
\author[4]{Xiangping Chen}
\author[2]{Shenghua Gao}
\author[1,3]{Jiang Liu}

\address[1]{\orgdiv{Cixi Institute of Biomedical Engineering}, \orgname{Ningbo Institute of Materials Technology and Engineering}, \orgaddress{\state{ Ningbo}, \country{China}}}

\address[2]{\orgdiv{School of Information Science and Technology}, \orgname{ShanghaiTech University}, \orgaddress{\state{Shanghai}, \country{China}}}

\address[3]{\orgdiv{Department of Computer Science and Engineering}, \orgname{Southern University of Science and Technology}, \orgaddress{\state{Shenzhen}, \country{China}}}

\address[4]{\orgdiv{The Electrical Engineering College}, \orgname{Guizhou University}, \orgaddress{\state{Guiyang}, \country{China}}}

\address[5]{Equally contribution}

\corres{*Jianlong Yang \email{yangjianlong@nimte.ac.cn}}


\abstract{Optical coherence tomography angiography (OCTA) requires high transverse sampling density for visualizing retinal and choroidal capillaries. Low transverse sampling causes resolution degradation, such as the angiograms in wide-field OCTA. In this paper, we propose to address this problem using deep learning. We conducted extensive experiments on converting the centrally cropped $3\times 3$ mm$^2$ field of view (FOV) of the $8\times 8$ mm$^2$ foveal OCTA images (a sampling density of 22.9 $\mu$m) to the native $3\times 3$ mm$^2$ \textit{en face} OCTA images (a sampling density of 12.2 $\mu$m). We employed a cycle-consistent adversarial network architecture in this conversion. The quantitative analysis using the perceptual similarity measures shows the generated OCTA images are closer to the native $3\times 3$ mm$^2$ scans. Besides, the results show the proposed method could also enhance signal-to-noise ratio. We further applied our method to enhance diseased cases and calculate vascular biomarkers, which demonstrates its generalization performance and clinical perspective.}



\maketitle


\section{Introduction}
In recent years, optical coherence tomography angiography (OCTA) has drawn tremendous attention in ophthalmology as a novel imaging modality to replace traditional fluorescein angiography (FA) and indocyanine green angiography (ICGA) \cite{Gao2016}. OCTA is non-invasive thus avoids the risk of dye. Compared with the overlapping depth information of vasculature in 2D imaging like FA and ICGA, OCTA is capable of resolving the vessels and capillaries in-depth direction with a high axial resolution of $5\sim 10$ $\mu$m. It has been widely used in studies of various ocular diseases, such as glaucoma \cite{liu2016}, age-related macular degradation \cite{JiaE2395}, and retinopathy of prematurity \cite{Campbell2017}. Also, the OCTA imaging of human retina was found to be able to indicate neurodegenerative disorders, such as mild cognitive impairment \cite{hong2018} and Alzheimer's disease \cite{Bryhim2018}.\\
\indent OCTA employs the temporal decorrelation of repeated B-scans at the same location to distinguish the blood flow from surrounding tissue. Two mainstream OCTA algorithms, optical micro-angiography (OMAG) \cite{Wang2010} and split spectrum amplitude-decorrelation angiography (SSADA) \cite{Jia2012}, require to repeatedly scan four and two times, respectively. Besides, for minimizing motion noise and artifacts and enhancing the signal-to-noise ratio (SNR) of blood vessels, hardware-based eye-tracking \cite{Braaf2012} and software-based orthogonal registration \cite{Camino2016} further increase the acquisition duration time of OCTA. So for a given field of view (FOV) and A-line rate, OCTA scans spend a much longer time than volumetric OCT scans.\\
\indent On the other hand, different from OCT that emphasizes the structure information of retinal layers, an important function of OCTA is to visualize \textit{en face} retinal and choroidal capillaries. So OCTA has higher demands on the transverse resolution and sampling density of the acquisition systems. However, the imaging of the posterior segment is suffered from the intrinsic aberrations of the eye, which limits the transverse resolution of the ocular imaging systems to $10\sim 15$ $\mu$m. Incorporating with adaptive optics (AO) could compensate the aberrations and improve the transverse resolution \cite{Salas:17}, but the AO components will significantly increase the cost and complexity of the acquisition systems. Besides, the resolution is improving in sacrifice of the FOV, which hinders the clinical applications of the AO OCTA.\\
\indent Enabling a high transverse sampling could avoid the loss of capillary details, namely, the degradation of digital resolution induced by undersampling. However, it will further increase the acquisition time on the basis of the repeated scans. To avoid the discomfort of the imaging subject caused by a long acquisition duration (the human spontaneous eye blink rate is $\sim15$ blinks per minute \cite{Doughty2006}), researchers and manufacturers are trending to use high-speed systems for the OCTA imaging. In research fields, the acquisition A-line rate has reached several megahertz \cite{Klein2017}. For commercial systems, 100-kHz swept-source (SS) OCT systems have emerged. However, the popularization of SS-OCT systems was significantly impeded by the limited wavelength regime and availability of sources with a broad bandwidth at the moment.\\
\begin{figure}[t!]
\centering\includegraphics[width=8cm]{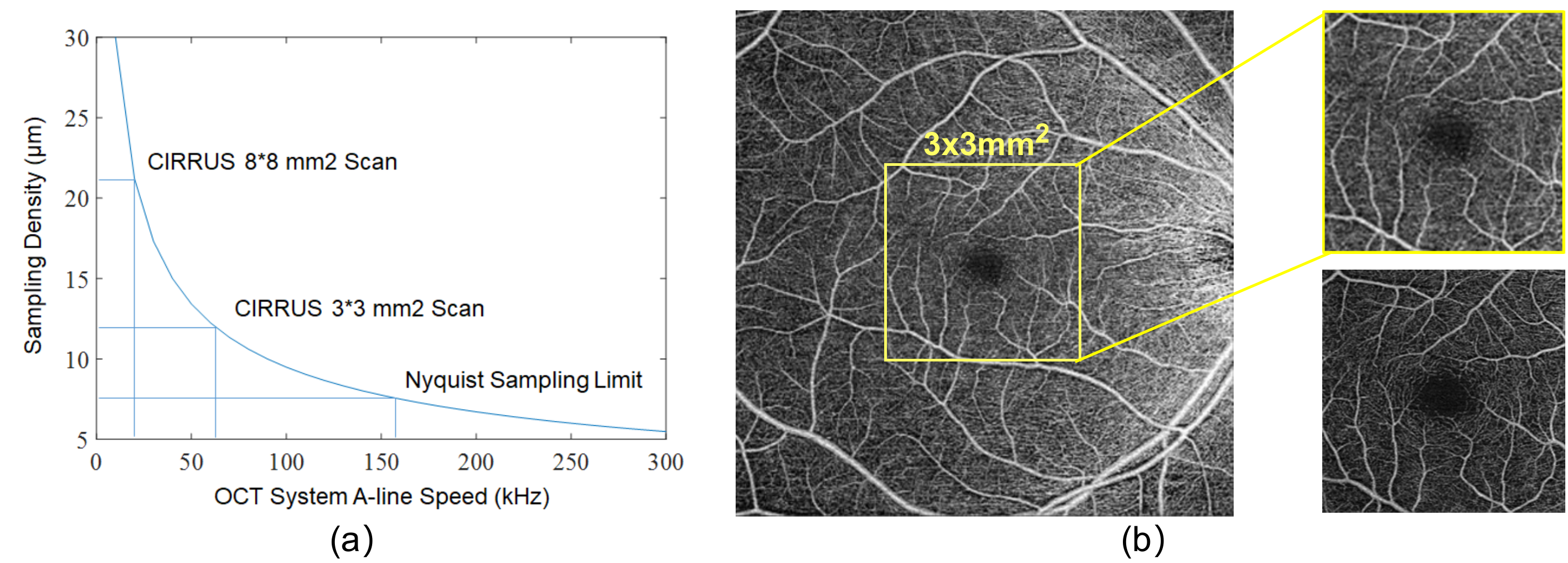}
\caption{(a) Relationship between the sampling density (rate) and the A-line speed of the OCT systems. (b) The comparison of $8\times 8$ mm$^2$ and $3\times 3$ mm$^2$ FOV OCTA images.}
\label{fig1}
\end{figure}
\indent Figure.~\ref{fig1}(a) give a simple estimation of the relationship between the sampling density and the A-line speed of the OCT systems. It was derived based on several practical assumptions. (1) An acquisition duration of 4 seconds, which is the upper limit time between two successive spontaneous blinks. (2) A FOV of $3\times 3$ mm$^2$, which is commonly used in OCTA scans. (3) Two repeated scans, which is the minimum for the OCTA algorithms. (4) Evenly-spaced sampling along the horizontal and vertical directions, which is mostly employed. For the sampling density at the Nyquist--Shannon limit (7.5 $\mu$m for an optical transverse resolution of $\sim15$ $\mu$m), an A-line speed of $\sim160$ kHz is required. We also listed the sampling density of the $3\times 3$ mm$^2$ and $8\times 8$ mm$^2$ scan protocols in the ZEISS CIRRUS OCTA system. For the  $3\times 3$ mm$^2$ scans, each transverse direction has 245 data points which corresponds to a sampling density of 12.2 $\mu$m. So the required A-line speed is $\sim 65$ kHz,  which is in accordance with the real-world situation. For the $8\times 8$ mm$^2$ scans, each direction has 350 data points which corresponds to a sampling density of 22.9 $\mu$m. So the demand for the A-line rate is reduced to $\sim 20$ kHz. However, insufficient sampling causes the blur and loss of the capillaries as demonstrated in Fig.~\ref{fig1}(b). The left side of Fig.~\ref{fig1}(b) shows the $8\times 8$ mm$^2$ superficial vascular plexus (SVP) scan centered on the fovea, the zoom-in view of its central FOV is shown in the upper left of the figure. The lower left of Fig.~\ref{fig1}(b) demonstrates the native $3\times 3$ mm$^2$ SVP image. 
\\ 
\indent The insufficient sampling not only affects the visualization and qualitative analysis of the OCTA images, but also brings difficulty and inaccuracy to the extraction of quantitative vascular biomarkers, such as vessel area density, vessel diameter index, and vessel perimeter index. Several studies have reported the discrepancy among the OCTA scan patterns with different FOVs \cite{Lei2017, Al-Sheikh2017,Li2018}. Usually, a large FOV scan would cause the underestimation of the vascular biomarkers mentioned above.\\
\begin{figure}[t!]
\centering\includegraphics[width=8cm]{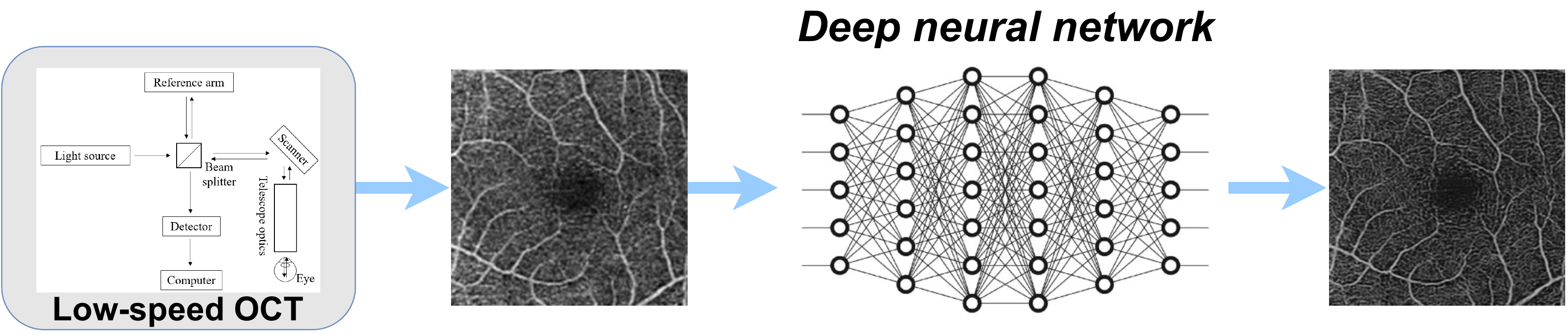}
\caption{Proposed workflow of using low-cost OCT systems and deep learning to get high digital resolution OCTA images.}
\label{fig2}
\end{figure}
\indent Here we propose to use deep learning techniques to remedy the digital resolution degradation caused the low transverse sampling in OCTA. We expect that it will have two major functionalities. (1) To improve the accuracy of extracting the vascular biomarkers using wide-field scans and/or low sampling scans. (2) Lower the speed requirement of the OCTA acquisition thus reduce the cost of the OCTA systems, which will promote the popularization of this technique. We illustrated this idea in Fig.~2. A low speed thus low-cost OCT system is used to acquire the OCTA data with a relatively low sampling density. Then the OCTA images are feed into a deep neural network and generates high-digital-resolution OCTA images. We believe this solution will benefit the development of healthcare in low-resource settings.
\\
\indent The idea of using deep learning in this low to high digital resolution conversion of OCTA is inspired by the success of the deep learning-based methods in enhancing the resolution of various types of microscopy very recently. Wang \textit{et al.} realized the transformation of ordinary fluorescence microscopic images into super-resolution images using a generative adversarial network (GAN) \cite{Wang2019}. Rivenson \textit{et al.} used a convolutional neural network to convert low-quality mobile-phone-based microscopic images to high-quality table-top microscopic images \cite{Rivenson2018}. de Haan \textit{et al.} demonstrated the conversion from low to high resolution scanning electron microscopic images using deep learning \cite{dehaan2019}.\\
\indent In this paper, we conducted extensive experiments to evaluate the feasibility of the proposed idea. we tried to convert the centrally cropped $3\times 3$ mm$^2$ field of view (FOV) of the $8\times 8$ mm$^2$ foveal OCTA images to the native $3\times 3$ mm$^2$ \textit{en face} OCTA images. A cycle-consistent GAN architecture was employed in this conversion. The results demonstrate the proposed deep-learning-based method is capable of significantly improving the digital resolution of the low transverse sampling OCTA images.
\section{Materials and methods}
\subsection{Data preparation}
40 eyes from 20 healthy participants (10 males and 10 females; ages ranging from 20 to 35 years old) were recruited for this digital-resolution-enhancement task at our Intelligent Ophthalmic Imaging \& Laser Laboratory. The human study protocol was approved by the Institutional Review Board of Cixi Institute of Biomedical Engineering, Chinese Academy of Sciences and followed the tenets of the Declaration of Helsinki. We employed a ZEISS CIRRUS OCTA system for the data collection. Each eye was scanned by two different imaging protocols: $3\times 3$ mm$^2$ and $8\times 8$ mm$^2$ FOV centered on fovea. The $3\times 3$ mm$^2$ and $8\times 8$ mm$^2$ scans have equivalent samplings of 245 and 350 along the two transverse directions, respectively. Each A-line has 1024 data points. The en face angiograms including the SVP, deep capillary plexus (DCP), choriocapillaris (CC), and choroidal vasculature were exported from the ZEISS acquisition and analysis software. We manually checked and corrected the results of automatic retinal layer segmentation. The low transverse sampling image was obtained by cropped $3\times 3$ mm$^2$ are centered on the fovea from wide-field image of $8\times 8$ mm$^2$ protocols. We defined the cropped $3\times 3$ mm$^2$ images as the original images and the GAN-processed images as the generated images below for simplicity. Also, the native $3\times 3$ mm$^2$ OCTA images are referred as high-definition (HD) OCTA images.
\\
\subsection{Deep learning network}
It's difficult to collect the paired OCTA images because of the deviation of scanning range adjusted by operators in each acquisition and the influences caused by the motions of living eyes. So the cycle-consistent adversarial network architecture \cite{Zhu17} can be used to repair the digital resolution degradation induced by the low transverse sampling with the unpaired images. The overall framework is illustrated in Fig.~3. It aims to learn a mapping $G_{AB}$ (mapping: A to B) such that the images from cropped $3\times 3$ mm$^2$ protocol is indistinguishable from the original $3\times 3$ mm$^2$ at both the pixel-level and feature-level with the adversarial learning. Then we couple it with an inverse mapping $G_{BA}$ (mapping: B to A) and introduce a cycle consistency loss to enforce the generated image is as similar to the reconstructed image as possible (and vice versa). Such a translation does not ensure that the input image and output image are paired up in a meaningful way, so two discriminator network $D_{A}$ and $D_{B}$ are brought out to restraint input images and synthesized images from generators in the same distribution.\\
\begin{figure}[htb]
\centering\includegraphics[width=8cm]{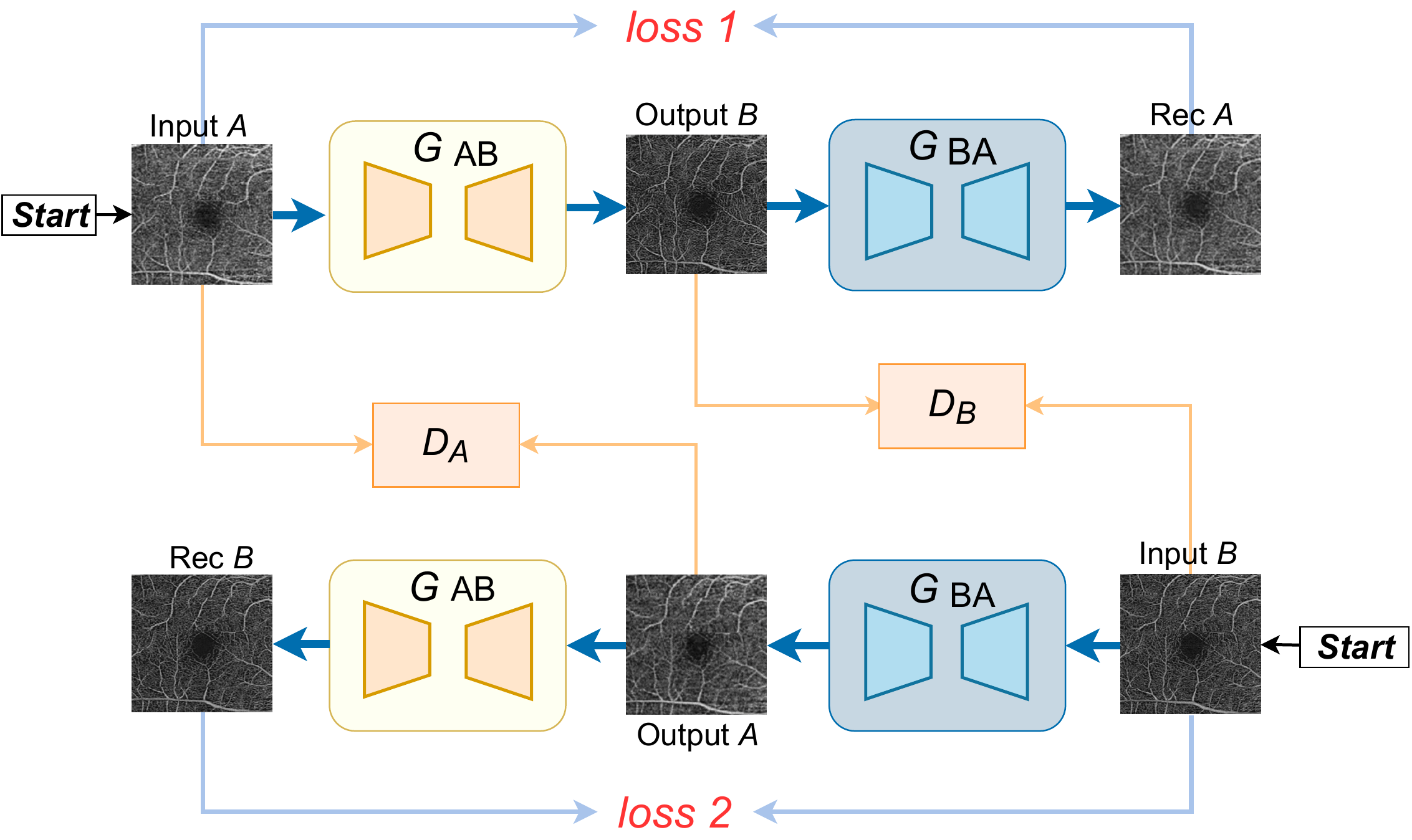}
\caption{Overview of the framework for the digital resolution enhancement task, there are two generators $G_{AB}$ and $G_{BA}$ and two discriminators $D_{A}$ and $D_{B}$. A denotes the low transverse sampling image domain, B denotes the high-definition image domain, to further regularize the mappings, the network employed three losses: cycle consistency loss, adversarial loss, and identity loss.}
\label{fig3}
\end{figure}
\indent  The network aims to learn the mapping between low-quality images (cropped $3\times 3$ mm$^2$ OCTA) and high-definition images (native $3\times 3$ mm$^2$ OCTA) and convert low-quality input images into high-quality output images. The overall loss is defined by:\\
\begin{equation}
\begin{aligned} L\left(G_{A B}, G_{B A}, D_{A}, D_{B}\right) &=L_{G A N}\left(G_{A B}, D_{B}, A, B\right) \\ &+L_{G A N}\left(G_{B A}, D_{A}, B, A\right) \\& +\beta L_{c y c}\left(G_{A B}, G_{B A}\right)\\ &+\gamma L_{i d e}\left(G_{A B}, G_{B A}\right) \end{aligned}
\end{equation}
\\
where $\beta$ and $\gamma$ control the proportion of each loss, $\ L_{G A N}$, $\ L_{c y c}$, $\ L_{i d e}$ denote the adversarial loss, cycle consistency loss, and identity loss. A denotes the low quality images (cropped $3\times 3$ mm$^2$ OCTA) in source domain, B denotes the high-definition images (native $3\times 3$ mm$^2$ OCTA) in target domain.\\
\indent Adversarial loss \cite{Goodfellow2014Generative} is applied in the mapping function, the objective is expressed as:\\
\begin{equation}
\begin{aligned} \min _{G_{A B}} \max _{D_{B}} L_{G A N}\left(G_{A B},\right.&\left.D_{B}, A, B\right) = E_{i_{b} \sim P_{d a t a}(b)}\left[\log D_{B}(b)\right]\\+&E_{i_{b} \sim P_{d a t a}(a)}\left[\log \left(1-D_{B}\left(G_{A B}(a)\right)\right)\right] \end{aligned}
\end{equation}
\\Where $G_{AB}$ is trained to enhance the digital resolution of low transverse sampling images to make it similar to the high-definition OCTA while $D_{B}$ aims to distinguish between images generated by $G_{AB}$ and real images in domain B. $G_{AB}$ aims to minimize this objective against an adversary $D_{B}$ that tries to maximize it. We employ a similar adversarial loss for the mapping function $G_{BA}$ and its discriminator $D_{A}$ as well.\\
\indent Adversarial training can learn mappings $G_{AB}$ and $G_{BA}$ that produce outputs identically distributed as target domain B and source domain A, respectively. However, the network may map the same set of input images to any image in the target domain, that is to say, if we only employ adversarial loss, it's hard to learn the function maps with a single input $A_{i}$ to the desired output $B_{i}$. So the cycle consistent loss is used to the network to constrain the mapping, for each image A from source domain, the image translation cycle should be able to bring A back to the original image:\\
\begin{equation}
\begin{aligned} L_{c y c}\left(G_{A B}, G_{B A}\right)&= E_{i a} \sim P_{d a t a}\left[\left\|G_{B A}\left(G_{A B}\left(i_{a}\right)\right)-i_{a}\right\|_{1}\right] \\ &+E_{i_{b} \sim P_{d a t a}(b)}\left[\left\|G_{A B}\left(G_{B A}\left(i_{b}\right)\right)-i_{b}\right\|_{1}\right] 
\end{aligned}
\end{equation}
\\Where the cycle consistency loss enforces the constraint that $G_{AB}$ and $G_{BA}$ should be inverse of each other, i.e., it encourages\\
\begin{equation}
G_{B A}\left(G_{A B}\left(i_{A}\right)\right) \approx i_{A} \text { and }G_{A B}\left(G_{B A}\left(i_{B}\right)\right) \approx i_{B}
\end{equation}
\indent In order to prevent excessive color composition between input and output, we regularize the input of network provide by target domain to be near an identity mapping:\\
\begin{equation}
\begin{aligned} L_{i d e}\left(G_{A B}, G_{B A}\right) = E_{i_{b} \sim P_{d a t a}(b)}\left[\left\|G_{A B}(b)-i_{b}\right\|_{1}\right]\\ +  E_{i_{a} \sim P_{d a t a}(a)}\left[\left\|G_{B A}\left(i_{a}\right)-i_{a}\right\|_{1}\right] \end{aligned} 
\end{equation}
\subsection{Implementation}
Our model is implemented based on the deep learning framework PyTorch in the Ubuntu 16.04 LTS operating system and the training of the model was performed with NVIDIA GeForce GTX 1080 Ti GPU with 12 GB RAM. We employed the open-source implementation of the cycle-consistent adversarial network\footnote{\url{https://github.com/junyanz/pytorch-CycleGAN-and-pix2pix}} shared by its inventors \cite{Zhu17}. The generator and discriminator are trained from scratch using the Adam optimizer \cite{kingma2014adam} with an initial learning rate of $2\times10^{-4}$ and a batch size of 1. We set $\beta_1=0.5$ and $\beta_2=0.999$ for both of the two Adam optimizers. In the training of the GAN model, the loss curve was shown to be convergent after 180 epochs, so we stopped at 200 epochs for achieving the smallest loss. It took about 4 hours for each training.\\
\indent In this study, 40 OCTA data sets were split into 35 training sets and 5 testing sets. We did not set up validation sets because of the very limited size of data sets. This splitting strategy has been commonly used both in medical image analysis and computer vision fields\cite{staal2004ridge, liu2018future, kermany2018identifying}. Besides, as mentioned above, the model converged well using this strategy, so the validation data sets may not be necessary. We also used the data augmentation methods for enhancing the training data sets, including Gaussian blurring, contrast adjustment, adding Gaussian noise, and random rotation. So a total of 280 data sets were used in each training. Then the model was tested on the 5 testing data sets and other diseased data sets. For each of the SVP, DCP, CC, and choroid, we trained their model separately following the same procedure described above.
\subsection{Quantitative evaluation metrics}
The vasculature captured the OCTA imaging of the exactly same region using the $8\times 8$ mm$^2$ and $3\times 3$ mm$^2$ scan protocols does not have the same spatial mapping (as shown in Fig.~1(b)), which is the primary obstacle for the quantitative evaluation of the results in this work. The discrepancy in the spatial mapping has also been observed in several OCTA repeatability and reproducibility studies \cite{hong2019intra,lei2017repeatability,men2017repeatability,lee2019repeatability}, which may primarily relate to eye motion, signal strength difference caused by focusing or alignment, undersampling alias, and segmentation inaccuracy. So we could not use the native $3\times 3$ mm$^2$ scans as the ground truth for calculating the pixel-wise image similarity metrics, such as SSIM, PSNR, Dice coefficient, and Jaccard index (IoU).
\subsubsection{Perceptual similarity measures}
Instead, we turned to employ the quantitative metrics that are used in the evaluation of GAN models. The GAN- synthetic images also do not have a paired ground truth, which is similar to the situation in this work. Inception score (IS) and Fr\'{e}chet inception distance (FID) are the two most widely-used metrics in measuring the performance of GAN models. However, the IS may not suit our task because it calculates the KL divergence between the properties of the generated images and the features of the objects among the ImageNet classes \cite{borji2019pros}. On the other hand, the FID calculates the Wasserstein-2 distance between the features of two images. It could be used here to measure which one between the original image and the generated images is more similar to the paired HD OCTA image. Because the FID are consistent with human perception \cite{heusel2017gans}, we refer to as the perceptual similarity here corresponding to the pixel-wise similarity.\\
\indent Using a single metric may lead to bias in the evaluation, so we employed the kernel inception distance (KID) as supplementary, which was proposed very recently \cite{binkowski2018demystifying} and becoming more and more popular in the studies of GAN models. Compared with the FID, the KID measure the squared mean discrepancy between features instead of the Wasserstein-2 distance, which has an unbiased estimator with a cubic kernel while the FID may be empirically biased \cite{li2019linestofacephoto}.\\
\indent We employed an open-source implementation of the FID and KID\footnote[1]{\url{https://github.com/abdulfatir/gan-metrics-pytorch}}. Both the FID and KID use the Inception v3 network for feature extraction then employ the extracted features to calculated the FID and KID scores. Lower FID and KID scores mean higher perceptual similarity. In the implementation, we could select the dimensionality of features including 64 (first max pooling features), 192 (second max pooling features), 768 (pre-aux classifier features), and 2048 (final average pooling features). Because different levels of features represent different aspects of perception, e.g., texture and structure, we included all of these feature dimensions in the evaluation.
\subsubsection{Vessel caliber change and SNR metrics}
\begin{figure*}[t!]
\centering\includegraphics[width=12cm]{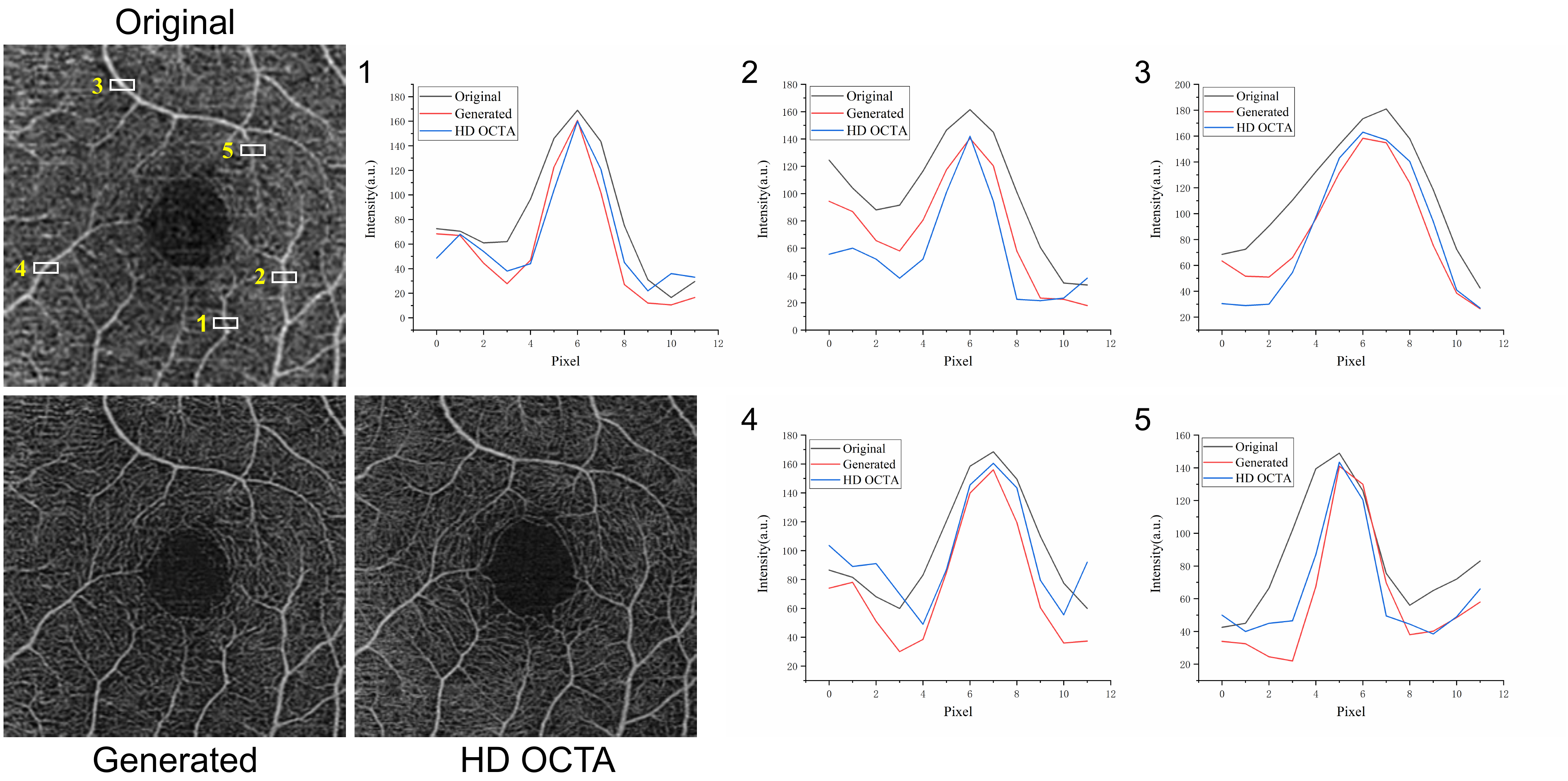}
\caption{Quantifying the changes of retinal vessel caliber. Left: 5 randomly selected positions of the vessels. Right: their intensity profiles. The black, red, and blue ones are from the original, generated, and HD OCTA images, respectively. } 
\label{vessl_fig}
\end{figure*}
To further characterize the performance of the proposed method, we quantified the changes of the vessel caliber and flow SNR before and after the digital resolution enhancement processing. These parameters of the HD OCTA are used as the reference.\\

\indent For calculating the changes of vessel caliber, we randomly selected 5 vessel positions in the SVP images of each test data set as demonstrated in the left side of Fig~\ref{vessl_fig}. Then we plotted their corresponding intensity profiles as shown in the right side of Fig.~\ref{vessl_fig}. We can see the vessel calibers are significantly decreased in the generated and HD OCTA image. To quantify these changes, we defined the discrepancy $S$ of the vessel calibers between the HD OCTA and the original or generated OCTA as
\begin{equation}
S = \frac{|W_{O, G}-W_{HD}|}{W_{HD}}\times 100\%,
\end{equation}
where $W$ is the full width at half maximum (FWHM) width of the intensity profile. $O$ and $G$ refer to the original and generated images, respectively. We calculated this value for all of the intensity profiles from each test data set and using their average and standard deviation as the quantitative measure of the vessel caliber changes. Because the noise is hard to differentiate with the vessels in the DCP, CC, and choroid, we only used the SVP for the evaluation here.
\indent We followed the calculation of the SNR in \cite{camino2017regression} as
\begin{equation}
SNR = \frac{\overline{D}_{parafoveal}-\overline{D}_{FAZ}}{\sigma_{D_{FAZ}}},
\end{equation}
where $\overline{D}_{parafoveal}$ is the averaged flow signal in the parafovea, which is concentric with the fovea, has an outer diameter of 2.5 mm and an inner diameter of 0.6 mm. $\overline{D}_{FAZ}$ is the averaged flow signal in the FAZ and $\sigma_{D_{FAZ}}$ is its standard deviation.
\subsection{Quantitative vascular biomarkers}
Quantitative assessment of OCTA is clinically useful for various ocular diseases\cite{Chu2016Quantitative}. We further applied the proposed method in the application of quantifying OCTA biomarkers. Following the methodology in \cite{Chu2016Quantitative}, we employed the following vascular biomarkers including vessel area density (VAD), vessel skeleton density (VSD), vessel diameter index (VDI), vessel perimeter index (VPI), vessel complexity index (VCI). These parameters have proven to be effective in the studies of ophthalmic diseases \cite{camino2016boe,alam2017quantitative,chen2017optical}.\\
\indent Because the DCP, CC, and choroid are severely contaminated by motion noises, we only calculated the vascular biomarkers of the SVP images. We first applied a hard threshold to remove the noises, which is two standard deviations above the mean value in the FAZ as defined in \cite{Jia2012}. Then we applied the 2D Frangi filter to the thresholded image with a scale range of [0.8 1.5], a scale ratio of 0.1. The Frangi $\beta_1$ and $\beta_2$ were set as 0.5 and 30, respectively. The vessel area map could further be acquired by using the MATLAB function \textit{imbinarize} with the adaptive threshold decided by the function \textit{adaptthresh}. Based on the vessel area map, we could calculate the vessel skeleton map and vessel perimeter map by using the functions \textit{bwskel} and \textit{bwperim}, respectively. Finally, the vascular biomarkers mentioned above could be easily calculated from the vessel maps by following the methods in \cite{Chu2016Quantitative}.
\section{Results}
\subsection{Visual comparison}
We first compare the images generated by the GAN model with the original input and the pair HD OCTA image visually.\\
\begin{figure}[h!]
\centering\includegraphics[width=8cm]{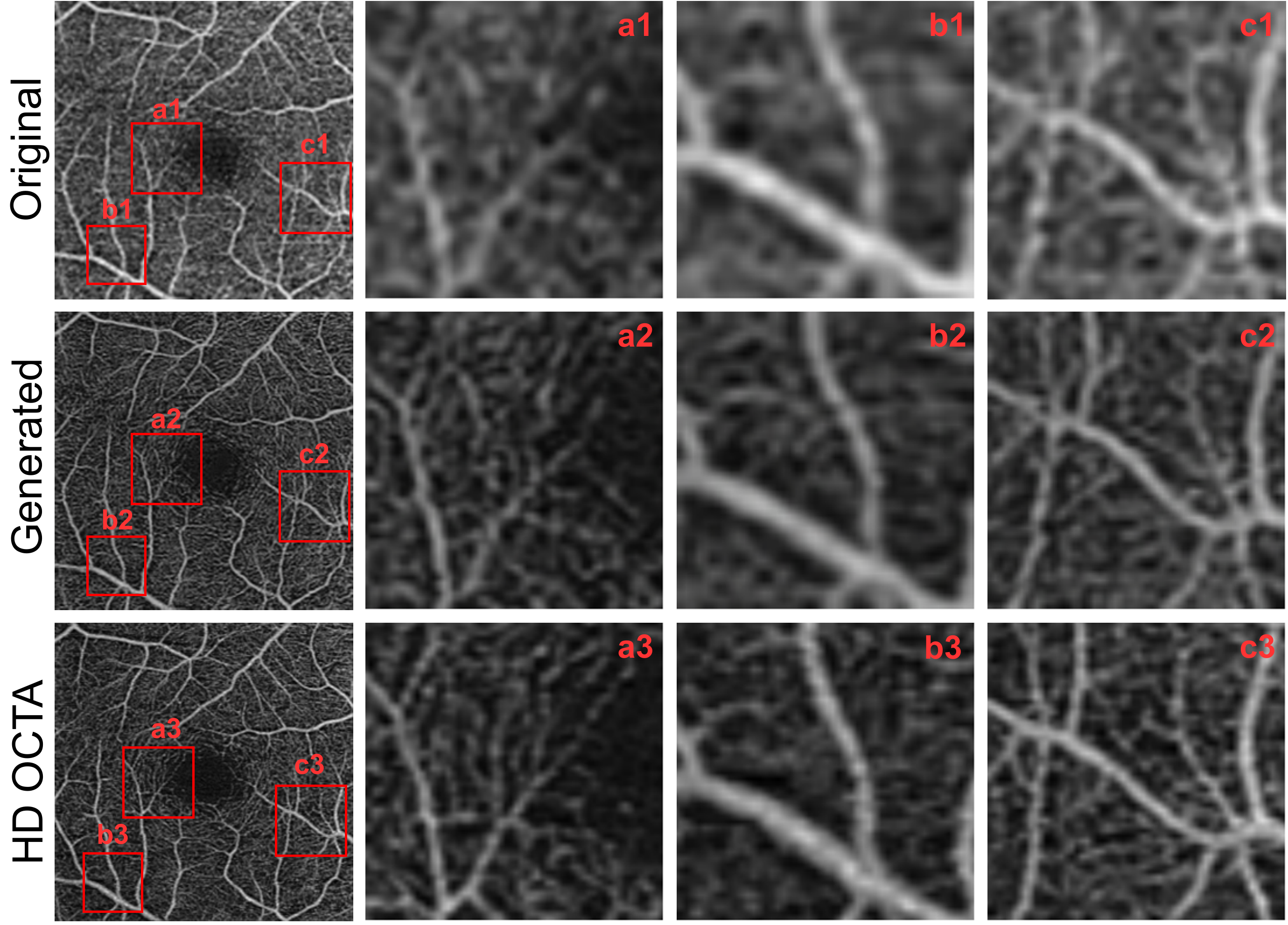}
\caption{Digital resolution enhancement results of the SVP, the zoom-in views of three regions of interest (ROIs) labeled as a, b, c (the red boxes) are demonstrated at the right side of the figure. The tag numbers 1, 2, and 3 refer to the original low transverse sampling image, the GAN-generated image, and the HD OCTA, respectively.} 
\label{fig4}\end{figure}
\indent Figure~\ref{fig4} shows the results of the SVP. The zoom-in views of three regions of interest (ROIs) labeled as a, b, c (the red boxes) are demonstrated at the right side of the figure. The tag numbers 1, 2, and 3 refer to the original low transverse sampling image, the GAN-generated image, and the HD OCTA, respectively. In the ROI (a), we can see the sharpness of the capillaries around the foveal avascular zone (FAZ) is significantly enhanced. Besides, the bulk motion noise inside the FAZ seems to be suppressed because of the conversion. Inside the ROI (b), large vessel branches are captured by the OCTA imaging, but the capillaries are almost absent even in the HD OCTA image. Because of the digital resolution enhancement, the calibers of the large vessels decrease in the generated image. The sizes of the vessels are similar to those in the HD image. The ROI (c) is located close to the edge of this $3\times 3$ mm$^2$ FOV. We can observe both the vessels and capillaries are in this region. The proposed method successfully converted the original blurry angiogram into the high-resolution image.\\
\begin{figure}[h!]
\centering\includegraphics[width=8cm]{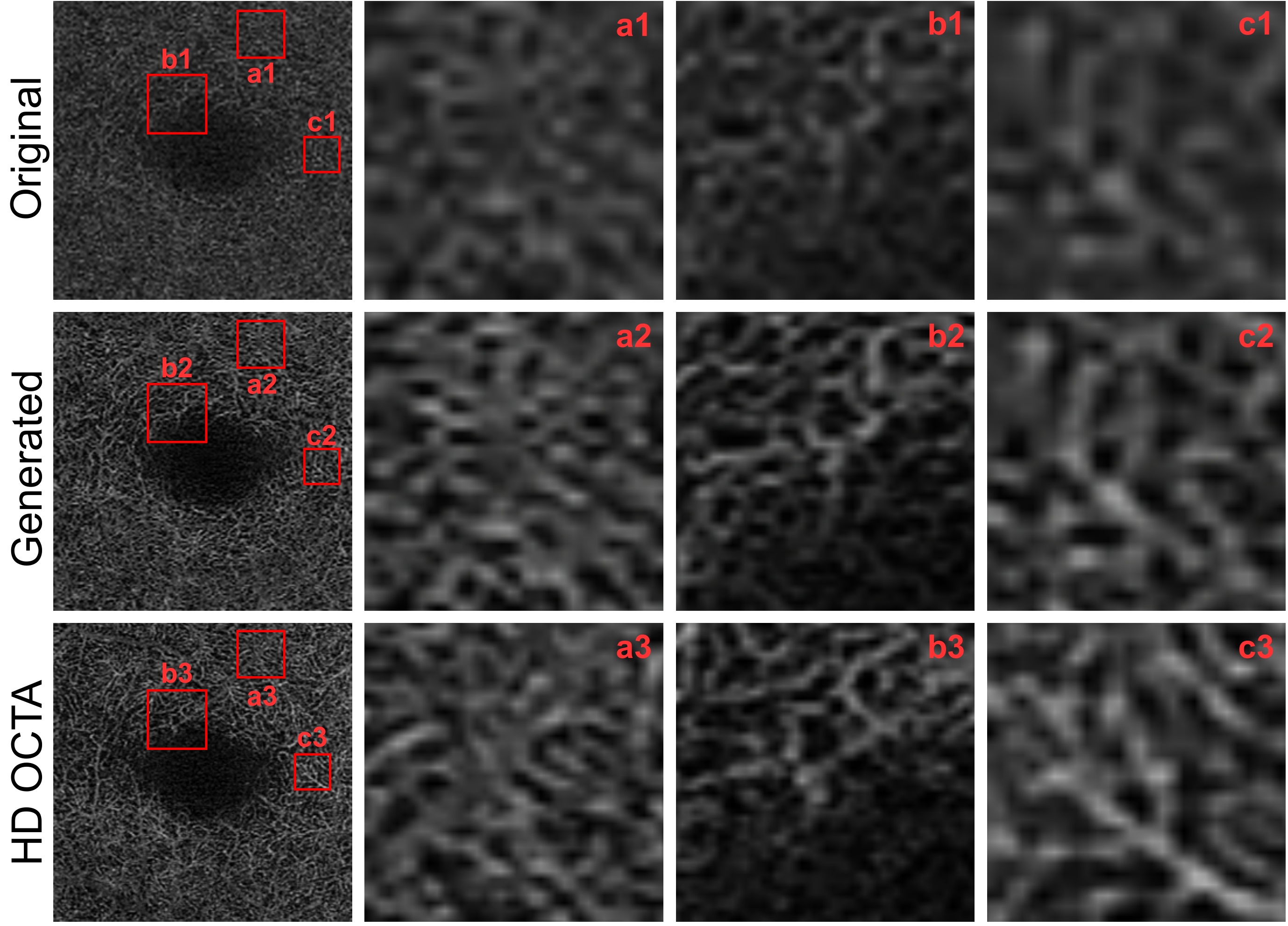}
\caption{Digital resolution enhancement results of the DCP. (a1-c1) The zoom-in views of interest (ROIs) labeled in the red boxes of original low transverse sampling image. (a2-c2) Demonstrated the corresponding zoom-in region of the GAN-generated image. (a3-c3)  Magnified corresponding area at the red boxes of HD OCTA.}
\label{fig5}\end{figure}
\indent The DCP is a purer and denser capillary plexus compared with the SVP as and we conducted a very similar analysis as demonstrated in Fig.~\ref{fig5}. The GAN-based conversion significantly improves the digital resolution and contrast of the original blurry images.\\
\begin{figure}[h!]
\centering\includegraphics[width=8cm]{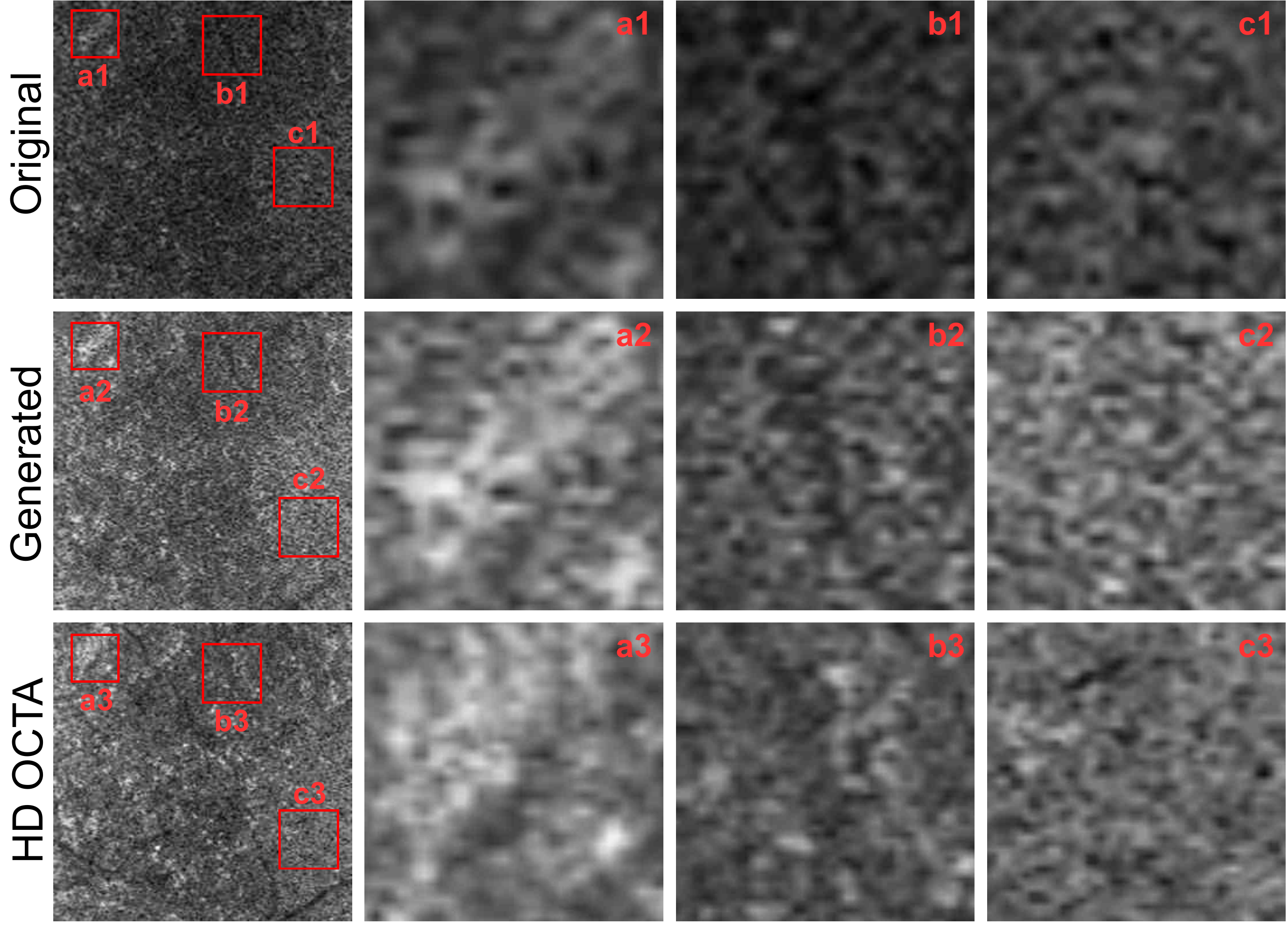}
\caption{Digital resolution enhancement results of the CC, the zoom-in views of three regions of interest (ROIs) labeled as a, b, c (the red boxes) are demonstrated at the right side of the figure. The tag numbers 1, 2, and 3 refer to the original low transverse sampling image, the GAN-generated image, and the HD OCTA, respectively.}
\label{fig6}\end{figure}
\begin{figure}[h!]
\centering\includegraphics[width=8cm]{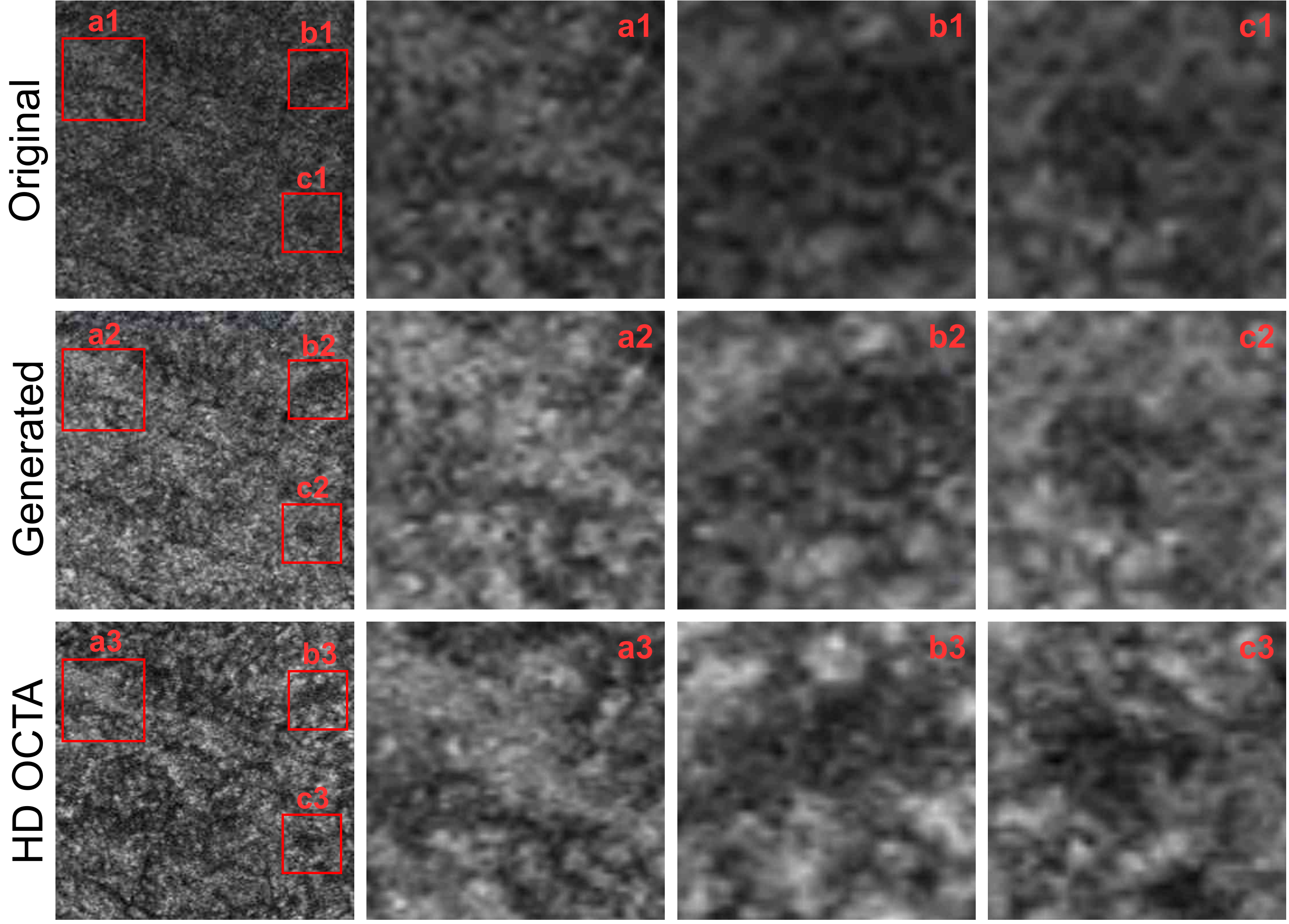}
\caption{Digital resolution enhancement results of the choroidal vascular, the zoom-in views of three regions of interest (ROIs) labeled as a, b, c (the red boxes) are demonstrated at the right side of the figure. The tag numbers 1, 2, and 3 refer to the original low transverse sampling image, the GAN-generated image, and the HD OCTA, respectively. }
\label{fig7}\end{figure}
\indent The results of the CC and the choroidal vasculature are demonstrated in Fig.~\ref{fig6} and \ref{fig7}, respectively. The contract of the CC and choroid images are significantly enhanced because of the GAN-conversion. The morphological features of the original images are well preserved. But it seems the proposed method does not improve the digital resolution of these two types of angiograms. \\
\indent As mentioned above, the original image and the HD OCTA do not have the same spatial mapping essentially, so we can observe that their capillary vasculatures have discrepancies with that of the HD OCTA image. Fortunately, the spatial mapping between the original and generated images is excellent, which means the proposed method does not create the features that do not exist.
\subsection{Quantitative evaluation}
Through the visual comparison above, we can see the proposed method only improves the contrast of the CC and choroid images and has minimal influence on the digital resolution, so we only include the SVP and DCP in the quantitative evaluation.

\subsubsection{Perceptual similarity}
As shown in Table 1, for both the SVP and DCP, the FID and KID scores of the generated images are lower than those of the corresponding original images which means they have better perceptual similarity with the HD OCTA images. It should be noted that the discrepancies between the scores of the original and generated images are larger in the lower feature dimensions (64 and 192), which is because the low-level feature maps are closer to the original images. while the high-level feature maps are more abstract.
\begin{table}[h!]
\centering
\caption{Quantitative comparison of the perceptual similarity measures}
\begin{tabular}{ccccc}
\hline
         & \multicolumn{2}{c}{SVP} & \multicolumn{2}{c}{DCP} \\
\hline
         & Original  & Generated  & Original  & Generated  \\
\hline
FID-64   & 9.101      & 0.445      & 11.138      & 0.392      \\
FID-192  & 24.521      & 1.664      & 18.381      & 1.471      \\
FID-768  & 0.815      & 0.328      & 1.322      & 0.715      \\
FID-2048 & 255.324      & 103.676      & 256.932     & 134.766      \\
KID-64   & 31.381      & 1.381      & 52.831     & 1.709      \\
KID-768  & 0.003      & 0.001      & 0.005      & 0.003      \\
KID-2048 & 0.527      & 0.225       & 0.497      & 0.278   \\  
\hline
\end{tabular}
\end{table}
\subsubsection{Vessel caliber change and SNR}
\begin{figure}[t!]
\centering\includegraphics[width=8cm]{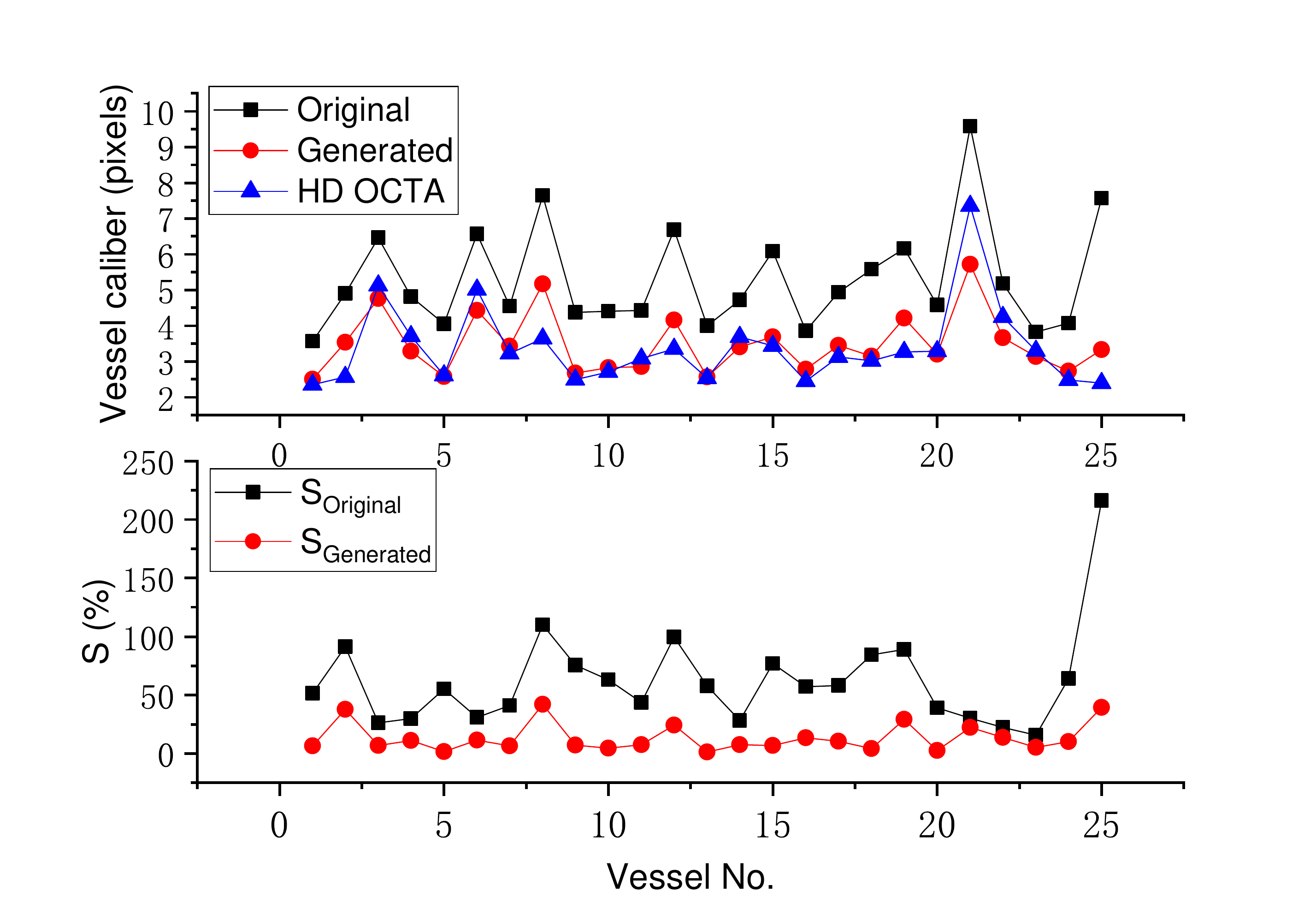}
\caption{Quantify the changes of vessel caliber. Top: the variation of vessel calibers in 25 random positions from the 5 testing data sets. The black, red, and blue lines are from the original, generated, and HD OCTA images, respectively. Bottom: The discrepancies between the original and HD (black) and the generated and HD (red).} 
\label{vesselCaliber}
\end{figure}
We plotted the variation of vessel calibers in 25 random positions from the 5 testing data sets in the top of Fig.~\ref{vesselCaliber}. The black, red, and blue lines are from the original, generated, and HD OCTA images, respectively. We also plotted the corresponding discrepancies between the original and HD (black) and the generated and HD (red) at the bottom of Fig.~\ref{vesselCaliber}. We can see the calibers of the vessel from the generated images are always smaller than those of the original images and closer to the HD images.\\
\begin{table}[h!]
\centering
\caption{Quantitative comparison of the SNR}
\begin{tabular}{cccc}
\hline
 & Original & Generated & HD OCTA\\
 \hline
 SNR &  $7.02\pm1.04$ & $9.96\pm1.94$ & $11.19\pm1.86$\\
 \hline
 \end{tabular}
\end{table}
\indent Table 2 is the quantitative comparison of the SNR. It shows the generated images have significant improvement in SNR, which is in accordance with the visual inspection above.
\subsection{Application in visualizing pathologies}
\begin{figure}[h!]
\centering\includegraphics[width=8cm]{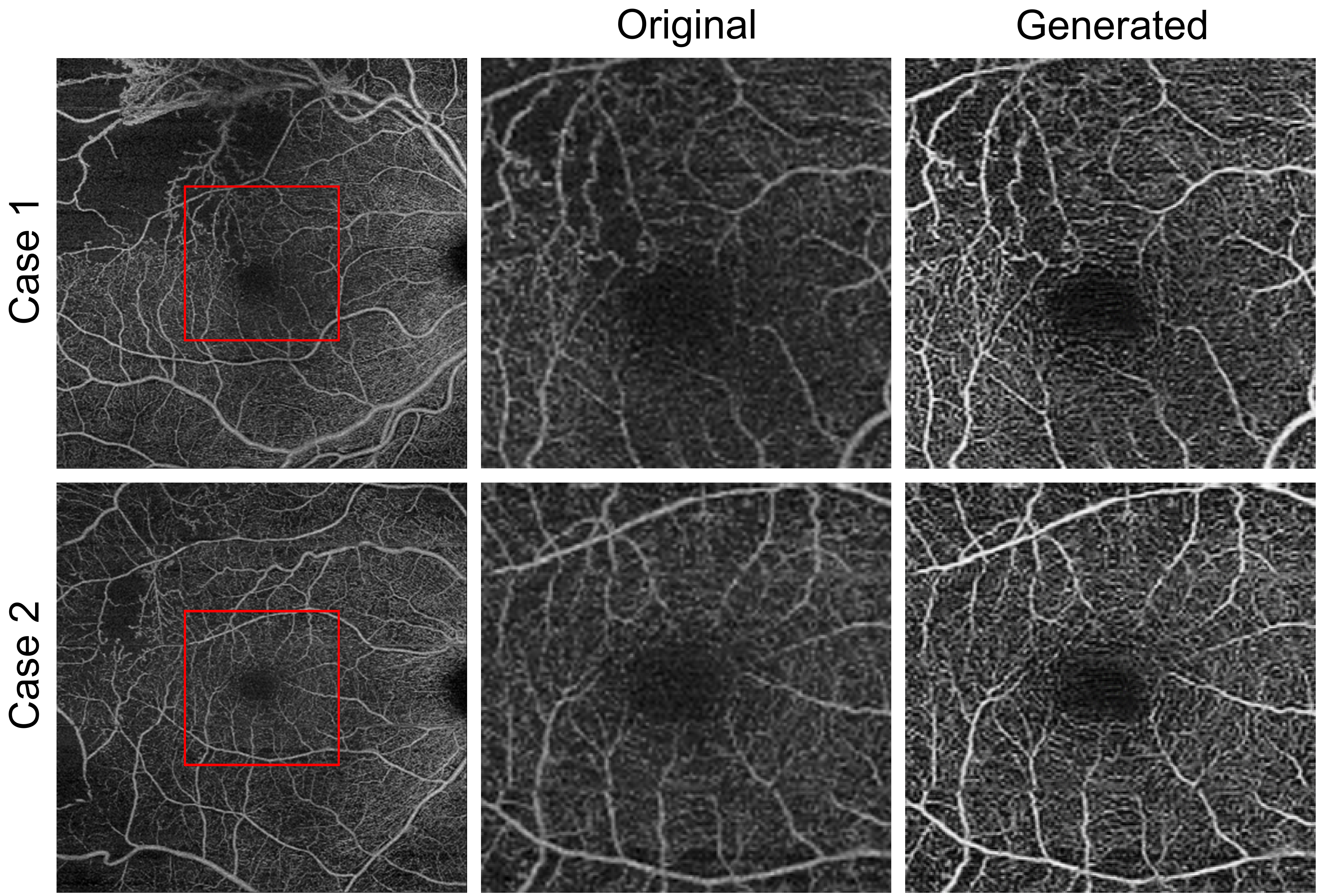}
\caption{Application of the proposed method in unseen diseased cases. The case 1 is from the right eye of a male (age 51), who has branch retinal vein occlusion. The case 2 is from the right eye of a 70-years-old subject (gender unknown), who has diabetic retinopathy. From left to right are the $8\times8$ mm $^2$ scans for these two cases, the cropped $3\times3$ mm $^2$ images referring to original, and the outputted images for the deep network referring to generated. The red boxes give the FOVs for the cropping.} 
\label{abnormal}
\end{figure}
For validating the generalization performance of the proposed method and evaluating its value in clinical applications, we tested our method on two unseen diseased cases. They are from the pre-installed database inside the ZEISS acquisition and analysis software. The case 1 is from the right eye of a male (age 51), who has branch retinal vein occlusion. The case 2 is from the right eye of a 70-years-old subject (gender unknown), who has diabetic retinopathy. Because the database only stores the $8\times8$ mm $^2$ scans for these two cases, so we do not have the HD OCTA images for comparison. The $3\times3$ mm $^2$ FOV images were cropped centered on the fovea as the input of the trained deep network.\\
\indent As demonstrated in Fig.~\ref{abnormal}, we can see the original cropped images have blurry vessels and low and unevenly distributed signal strength. After applying the proposed method, the vessels in the generated OCTA images are much sharper and narrower than those of the original images. Besides, we can see the visibility of the capillaries around the FAZ is significantly improved in the generated images. As mentioned above, our model was trained using normal images only, so its performance on the diseased data is an excellent indication of its potential in clinical applications.
\subsection{Application in quantifying vascular biomarkers}
To further validate the proposed method, we conducted a quantitative analysis of the SVP by comparing the vessel feature maps and their biomarkers from the original, GAN-generated, and HD OCTAs. Figure.~8 shows the images and maps of the SVP. (a) are the OCTA images. (b), (c), and (d) are the vessel area map, vessel skeleton map, and vessel perimeter map, respectively. We can see the vessel maps of the generated image and the HD OCTA are quite similar, while the vessel maps calculated from the original image are evidently sparser, especially at the regions away from the FAZ.\\
\begin{figure}[h!]
\centering\includegraphics[width=8cm]{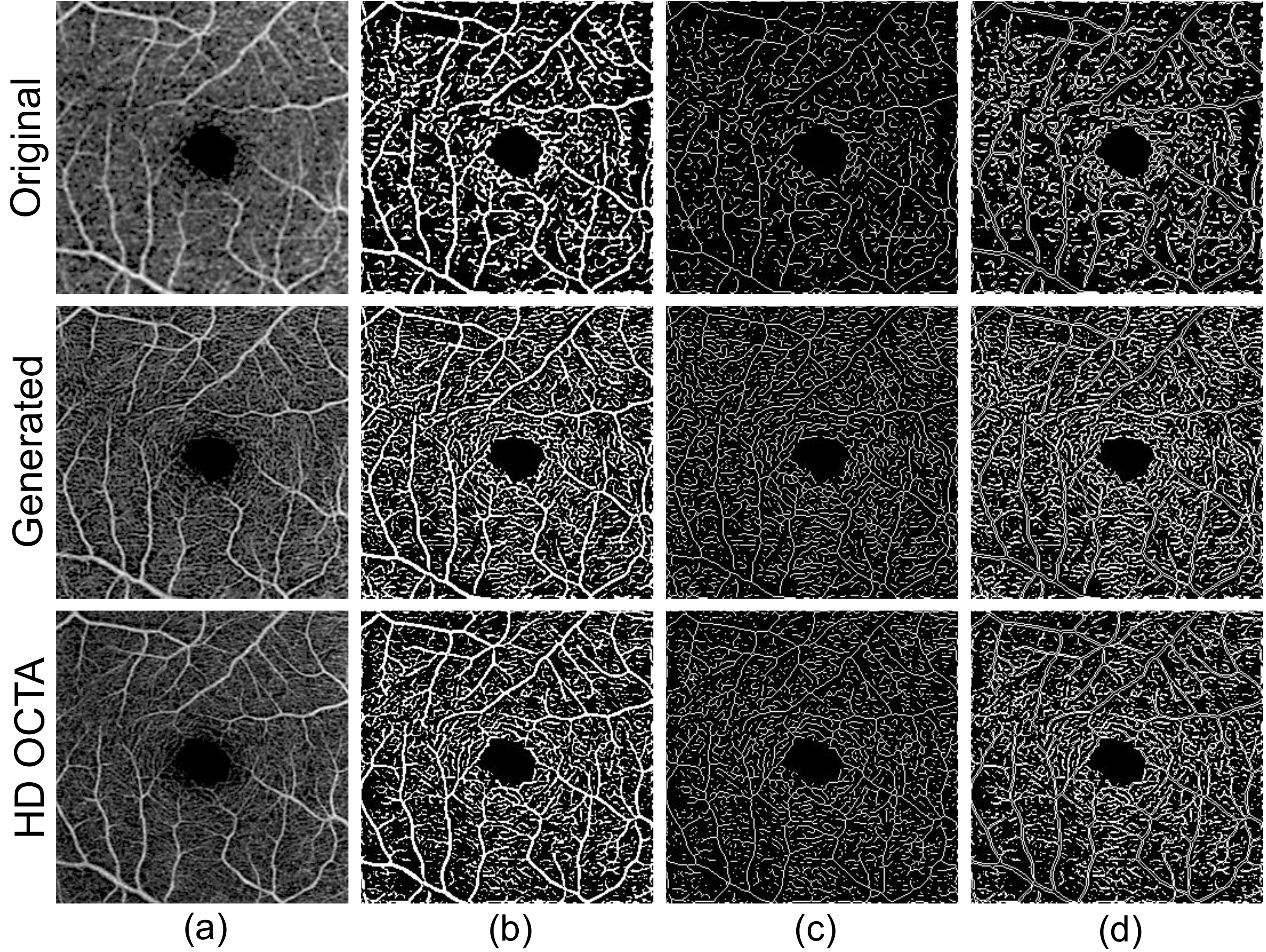}
\caption{Representative OMAG images to calculate the quantitative indexes of the SVP. (a) Original \textit{en face} OCTA image. (b) Vessel area map is used for VAD, VDI, and VCI quantification. (c) Vessel skeleton map is used for VSD and VDI quantification. (d) Vessel perimeter map is used for VPI and VCI quantification.}
\end{figure}\label{fig8}
\begin{table*}[t!]
\caption{Comparison of the vascular biomarkers of the SVP.}
\centering\begin{tabular}{@{}cccccc@{}}
\cline{1-6}
\multicolumn{1}{c}{}                                    & VDI         & VAD         & VSD         & VPI         & VCI               \\ 
\cline{1-6}
Original                                                & 2.514$\pm$0.027 & 0.261$\pm$0.009 & 0.104$\pm$0.004 & 0.204$\pm$0.008 & (1.314$\pm$0.055)$\times10^5$ \\
Generated                                               & 1.923$\pm$0.015 & 0.293$\pm$0.005 & 0.152$\pm$0.002 & 0.261$\pm$0.005 & (1.924$\pm$0.039)$\times10^5$ \\
HD OCTA                                                 & 1.868$\pm$0.027 & 0.282$\pm$0.016 & 0.151$\pm$0.010 & 0.250$\pm$0.017 & (1.818$\pm$0.141)$\times10^5$ \\
\cline{1-6}
\end{tabular}
\end{table*}
\indent Based on the vessel maps, we calculated the vascular biomarkers of these three types of images including the VDI, VAD, VSD, VPI, and VCI as listed in Table 3. The VDI is an indicator of the vessel diameters. We can see the generated SVP image has a low VDI compared with the original one, which confirms the improvement of the digital resolution. The VAD, VSD, VPI, and VCI are all indicators of the abundance of the capillaries from different aspects of the vasculature (length, caliber, and so on). The low transverse sampling will cause the underestimation of these biomarkers. The deep-learning-based conversion, on the other hand, can effectively minimize the underestimation. Besides, the vascular biomarkers extracted from the generated SVP image is in good accordance with those extracted from the HD SVP image. \\
\section{Discussion and conclusions}
Enhancing the digital resolution of the low transverse sampling data using deep learning loses the requirement of high-speed acquisition systems in OCTA, which has the potential to reduce the cost of clinical OCTA machines and promote the applications of this technique in a wider population. This paper demonstrates the implementation of this idea by training the low transverse sampling \textit{en face} OCTA images to learn the features of the high transverse sampling data using a cycle-consistent adversarial deep network. The results have demonstrated the proposed conversion could significantly improve the digital resolution without disrupting the original morphological features of the retinal vessels. Besides, we have found the improvement in digital resolution also benefited the accurate quantification of the vascular biomarkers, such as vessel area density and vessel perimeter index.\\
\indent However, OCTA is a 3D imaging modality, so the \textit{en face} projected images can not provide intact depth information. Also, this work only enhanced the digital resolution of the $3\times3$ mm$^2$ FOV centered on the fovea, which is far from sufficient for the researches and diagnoses of ocular diseases, especially the early symptoms happening at the peripheral retina, such as the early stages of diabetic retinopathy. In the next step, we will devote to the development of this new technique in two major directions. (1) The digital resolution enhancement of OCTA B-scans (volumetric OCTA). Because it will involve the axial direction, we will consider to introduce compressed sensing techniques to fulfill this task. (2) Enhancing the digital resolution in larger FOVs. For example, including fovea and optical nerve head simultaneously. We will combine the style transfer deep network with other techniques such as deep learning or low-level feature-based registration and impainting. (3) Not only the digital resolution, this deep-learning-based approach could also be used in optical resolution enhancement, by using AO OCTA data as the target domain. Because of the unpaired transition capacity, instead of using the paired OCTA images captured when the AO loop is turned off and on, we may only need some arbitrary AO OCTA images for training.\\
\indent In summary, We have developed a deep learning-based technique that could enhance the digital resolution of the low transverse sampling OCTA. Because it is difficult to collect the paired OCTA images due to the deviation of scanning range adjusted by operators in each acquisition and the influences caused by the motions of living eyes, we have employed the cycle-consistent adversarial network architecture for this task. Qualitative and quantitative results have demonstrated the proposed technique could not only improve the transverse digital resolution and SNR of the OCTA images but also benefit the quantification of the vascular biomarkers.

\section*{Funding}
Ningbo 3315 Innovation team grant; Cixi Institute of Biomedical Engineering, Chinese Academy of Sciences (Y60001RA01, Y80002RA01); Zhejiang Provincial Natural Science Foundation (LQ19H180001); Ningbo Public Welfare Science and Technology Project (2018C50049).

\section*{Acknowledgments}
We would like to thank the reviewers and editors for the careful reviewing and insightful comments, which significantly benefits the improvement of this manuscript and helps us refine this new research topic. We also acknowledge the contribution of Zhengjie Chai and Qiangjiang Hao in the calculation of the quantitative metrics FID and KID.

\section*{Disclosures}
The authors declare no conflicts of interest.

\bibliography{wileyNJD-ACS}%

\end{document}